\documentclass{aastex}
\usepackage{emulateapj5}
\begin{document}

\newcommand{\ergscm}{erg~s$^{-1}$~cm$^{-2}$}

\title{A VLT/FORS2 Multi-Slit Search for Ly-$\alpha$ Emitting Galaxies at $z\sim6.5^{1}$}

\footnotetext[1]{Based on observations obtained at the
  European Southern Observatory, Paranal, Chile (ESO programme
  71A-3071A).}


\author{Kim-Vy H. Tran$^{2}$}
\author{Simon J. Lilly}
\affil{Institute for Astronomy, ETH Z\"urich, CH-8093
  Z\"urich, Switzerland}
\author{David Crampton}
\affil{Herzberg Institute of Astrophysics, 5071 West Saanich Road,
Victoria, BC V9E 2E7, Canada}
\author{Mark Brodwin}
\affil{Jet Propulsion Laboratory, California Institute of Technology, 
Mail Stop 169-506, Pasadena, CA 91109}

\footnotetext[2]{vy@phys.ethz.ch}

\setcounter{footnote}{3}

\begin{abstract}
  
  We present results from a deep spectroscopic search in the 9150\AA~
  atmospheric window for $z\sim6.5$ Ly$\alpha$ emitting galaxies using
  the VLT/FORS2.  Our multi-slit+narrow-band filter survey covers a
  total spatial area of $17.6$~arcmin$^{2}$ in four different fields
  and reaches fluxes down to $5\times10^{-18}$~\ergscm~($7\sigma$
  detection).  Our detection limit is significantly fainter than
  narrow-band searches at this redshift and fainter also than the
  unlensed brightness of Hu et al.'s HCM6A at $z=6.56$, and thus
  provides better overlap with surveys at much lower redshifts.
  Eighty secure emission line galaxies are detected.  However, based
  on their clear continuum emission shortward of the line or the
  presence of multiple lines, none of these can be Ly$\alpha$ emission
  at $z\sim6.5$.  Our null result of finding no $z\sim6.5$ Ly$\alpha$
  emitters suggests that the number density of Ly$\alpha$ emitters
  with $L\geq2\times10^{42}$ erg~s$^{-1}$ declines by $\gtrsim2$
  between $z\sim3$ and $z\sim6.5$.

\end{abstract}


\section{Introduction}

Great progress has been made over the last five years in the search
for high redshift ($z>5$) galaxies.  Several groups have successfully
detected galaxies beyond $z\sim5$ using both the continuum-based
``drop-out'' approach and searches for strong emission lines using
either narrow-band filter imaging or dispersed long-slit spectroscopy
\citep{rhoads:03,kodaira:03,stanway:04,dickinson:04,lehnert:03,maier:03,hu:04}.
A few surveys also have targeted the caustics in lensing galaxy
clusters to take advantage of the associated flux amplification
\citep{ellis:01,santos:04,kneib:04}.

The relative advantages of these different survey approaches are
fairly clear.  Continuum selection is most closely tied to
star-formation rate but is in practice limited to relatively luminous
objects.  Furthermore, the establishment of a true continuum ``break''
or ``drop-out'' requires two filters longward of Ly$\alpha$, limiting
the applicability of the method to $z\leq5.5$ unless deep wide-field
infrared imaging data is available.  In comparison, narrow-band
Ly$\alpha$ based searches avoid this difficulty but obviously only
pick out objects with strong Ly$\alpha$ emission.  For galaxies with
high equivalent widths, Ly$\alpha$ surveys can reach to fainter
continuum levels than the continuum based techniques.  Relative to
narrow-band imaging, longslit dispersed spectroscopic surveys achieve
a large increase in sensitivity due to the reduced background, but at
the cost of severely reduced sky coverage.  The use of cluster lensing
amplification also enables much fainter objects to be detected, but
involves only a small survey area in the source plane
($\leq0.5$~arcmin$^2$).  Lensed samples also require careful analysis
to avoid biases introduced by the peculiarities of the lensing
(Porciani et al, in preparation).

An alternative approach developed by \citet[hereafter
CL99]{crampton:99} is to combine multiple parallel longslits with a
narrow-band filter that limits the observed spectral range to a few
hundred angstroms.  This technique is optimal for a targeted redshift
survey because we achieve the full sensitivity gain of a high
resolution spectrum and increase the spatial coverage by the number of
long-slits.  The redshift range is the same as that of a narrow-band
survey, but this allows optimum use of narrow atmospheric windows such
as that at $9000-9300$\AA.  However, unlike a narrow-band imaging
survey, there is no significant selection in emission line equivalent
width.  Furthermore, spectral line identification diagnostics such as
line asymmetries, line doublets etc., as well as precise redshifts are
obtained in the first pass.  In principle, this eliminates the need
for follow-up spectroscopy to confirm the reality and characteristics
of emission lines.  The multi-slit/filter approach has been used on
CFHT by CL99, and on Keck/LRIS by \citet{stockton:99} and
\citet[hereafter MS04]{martin:04} to search for Ly$\alpha$ emitters
(LAE).  However, none of these surveys detected any LAEs at $z>5$.

In this paper, we present the results from a search for $z\sim6.5$
galaxies using the multi-slit/filter approach in the {9150 \AA} window
on the VLT/FORS2.  The program was initially motivated by the first
galaxy at $z>6$ found by \citet{hu:02} within an $0.4$~arcmin$^2$ area
behind the lensing cluster A370.  Our own observations were designed
to reach below the de-amplified brightness of HCM6A and to sample an
area 50 times larger.  Compared to the initial multi-slit/filter study
by CL99, the current survey is about five times deeper and covers
twice the spatial area.  It also surveys a volume five times larger
than that of \citet{stockton:99} and MS04.  Our study is the first to
survey a relatively large volume at $z\sim6.5$ to the low luminosities
comparable to that of most galaxies in $z\sim3$ surveys.

Figure~\ref{eqw_line} illustrates some of the features of these
different approaches by comparing emission line searches at $z>5$ with
the large continuum selected sample at $<z>=3$ of \citet{shapley:03}.
Many of the Ly$\alpha$ selected objects have continuum luminosities
overlapping those of the continuum selected ``drop-out'' objects, but
the continuum levels can also extend well below this level for high
equivalent width galaxies, especially when the line flux detection
limit is itself reduced, as in the present work.  The dispersed surveys
have a much lower equivalent width limit than the narrow-band imaging
surveys. However, as can be seen on the figure, to fully exploit this
would require an even lower line flux limit because of the apparent
maximum continuum luminosity exhibited by high redshift galaxies.

We adopt $\Omega_M=0.3$, $\Omega_\Lambda=0.7$, and
H$_0=70$~km~s$^{-1}$~Mpc$^{-1}$ throughout the paper.

\section{Summary of Observations and Data}

\subsection{Re-analysis of the CFHT Data (CL99)} 

As part of the main VLT program, we have re-analyzed our 1999 CFHT
data (CL99). A total of 11 emission line objects with line fluxes of
$f\geq2\times10^{-17}$~\ergscm~at wavelengths of
$9000\leq\lambda\leq9250$\AA~were detected.  One emission line object
was identified as a candidate galaxy at $z\sim6.43$ because of its
high observed equivalent width ($W_{obs}\sim200$\AA), absence of
detected continuum shortward of the line, and the absence of an
identifiable counterpart in deep $UBVRI$ imaging.  However, follow-up
spectroscopy taken with the VLT/FORS2 in July 2003 showed that the
emission was in fact H$\alpha$ at $z\sim0.4$ associated with a
relatively bright galaxy that lay outside of the longslit aperture.
The most important lesson from this was the need to map contiguous
areas of the sky by stepping observations perpendicular to the
longslits, thus minimizing the confusing effects of neighboring
objects, and to obtain accurate astrometry along the slit.  In
addition, it emphasized that high observed equivalent widths
($W_{obs}\sim200$ \AA) can not be reliably associated with Ly$\alpha$,
as also stressed by \citet{stern:00} and \citet{rhoads:04}.

The CFHT survey thus resulted in no Ly$\alpha$ galaxies above
$f\geq2\times10^{-17}$~\ergscm~(i.e. $L>3L^{\ast}$, assuming
$L^{\ast}=3.2\times10^{42}$~erg~s$^{-1}$) in an area of 9~arcmin$^2$
with $\Delta z=0.2$ (equivalent to a volume of
$4170h_{70}^{-3}$~Mpc$^3$).

\subsection{New VLT/FORS2 Observations}

To extend this program, further observations were made with the
VLT/FORS2 spectrograph during 23--26 July 2003 of four fields: the
HDF-South \citep{williams:00}, the Chandra Deep Field
South/GOODS-S/UDF region \citep{giavalisco:04}, the CFRS-22 hour field
(CFRS22), and one of the CTIO Deep Lens Survey fields (DLS-F5).  In
each of these four widely spaced areas, our observations sparsely
sample a $6\times7$~arcmin$^2$ area, thus minimizing the effects of
cosmic variance on the survey statistics.  The $600z$ VPH grism gave
1.6\AA~per $2\times2$ binned pixel, each of which covered
$0.25\times0.25$ arcsec$^2$.  The FORS2 slit-mask had nine parallel
longslits, each $2''$ wide and $415''$ long, spaced $42''$ apart; the
same mask was used for all observations.  Each of the four fields were
observed with two or three adjacent mask pointings that were displaced
from each other by $1.5''$ perpendicular to the slit direction.  The
total spatial area covered by our FORS2 observations is thus
$17.6$~arcmin$^2$.  The total exposure times for each pointing was
$8100-10800$ seconds.

The seeing varied between $0.5-1.0$ arcsec with an average of
$\sim0.7''$, leading to a spectral resolution for point sources of
about $5$\AA, sufficient to identify the [OII]$\lambda3727$ doublet at
$z\sim1.4$ and to detect the line asymmetry often associated with
Ly$\alpha$ emission.  Wavelength and flux calibration were achieved
with an Argon arc and observations of the standard star LTT6248. The
data were reduced using a combination of IRAF and D.~Kelson's Python
package.

\subsection{Detection of Emission Lines}

The variable response of the grism over the CCD was removed using a
domeflat, and the data were flat-fielded using a master flat created
from all the science images where bright objects had been masked.  The
sky emission was removed by subtracting a running median through the
data at constant wavelength.  A global continuum subtraction was then
made by fitting a linear function of wavelength along each spatial
pixel along the slit.  For optimal emission line detection, the final
image was normalized by a noise-map that had been derived from the
statistics of the coadded frames (averaged over $25''$ along the slit).
This noise-map reflected the effects of the varying sky brightness
across the spectral window due to OH emission lines, the effects of the
filter transmission in wavelength and of the blaze angle of the
volume-phase holographic grating from slit to slit, as well as the
different exposure times both along the slit and from observation to
observation.  To exclude regions with unacceptably high noise levels,
we truncate the wavelength range to $9035<\lambda<9225$\AA.  Apart from
the residuals from particularly bright objects, this procedure resulted
in a two-dimensional image of very uniform appearance in which emission
lines appeared as point sources.

Emission lines were detected using SExtractor v2.2 \citep{bertin:96}.
Subsequent measurement of the line fluxes and uncertainties as well as
the equivalent widths of the lines were measured from the original data
using aperture ($3''$ diameter) photometry.  From visual inspection of
detected emission lines, we concluded that emission lines with formal
$S/N>7$ were robust.

The noise-map was used to determine the survey volume as a function of
the line flux detection limit.  The deepest part of the survey reaches
$5 \times10^{-18}$~\ergscm, approximately a half reaches a depth of
$6\times10^{-18}$~\ergscm, and the whole survey volume of
6130$h_{70}^{-3}$~Mpc$^{-3}$ is covered at
$1.8\times10^{-17}$~\ergscm; these values correspond to $7\sigma$
detection limits.

\subsection{Identification of Emission Lines}

A necessary condition for identification of an emission line as
Ly$\alpha$ at $z\sim6.5$ is that there be no detected continuum
shortward of the line.  To search for such emission, we used our own
spectra as well as the publicly available $bviz$ imaging of GOODS-S,
the existing public imaging of HDFS (both HST and ground-based)
supplemented with a new 3hr $R$-band image obtained with FORS2 during
our own observing run, our own $UBVIRZ$ for the CFRS22 field
\citet{brodwin:04}, and imaging in $BVRz$ for the DLS-F5 field kindly
made available by the DLS team.

The final SExtractor catalog of emission lines from all ten pointings
contained $\sim80$ objects. Most objects were in fact easily identified
as lower redshift galaxies from their FORS2 spectra alone, e.g. the
[SII]$\lambda6717,6731$ pair, the [OIII]$\lambda4959,5007$ pair, and
the resolved [OII]$\lambda3727$ doublet.  In all but one case the
Ly$\alpha$ identification is securely excluded based on continuum
emission extending to much shorter wavelengths in the deep imaging
data.  The final object in which no continuum was detected is
identified as [OIII]$\lambda5007$ based on a second weaker emission
line at the correct wavelength for [OIII]$\lambda4959$.

We thus have a secure null result: There are no Ly$\alpha$ emitting
galaxies within the volume and flux limits of our current survey.
Using Poisson statistics \citep{gehrels:86}, we can place a limit of
less than 1.84 objects at the 84\% (one-sided) confidence level.

\section{Discussion}

\begin{figure}
\epsscale{0.7}
\plotone{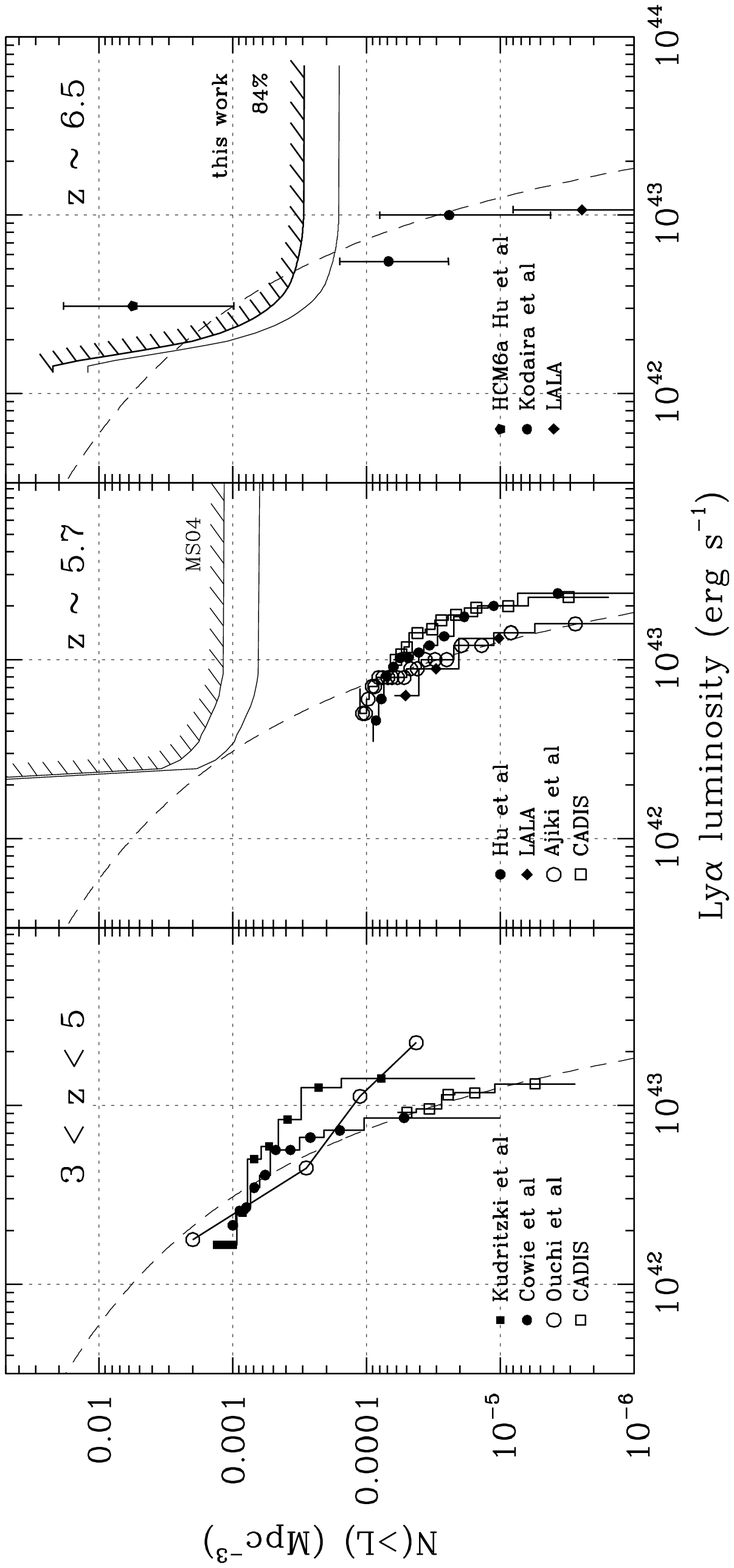}
\caption{The cumulative luminosity functions of Lyman-$\alpha$
  emitters at $3<z<5$ (left), $z\sim5.7$ (middle), and $z\sim6.5$
  (right).  Open symbols denote photometrically selected LAE samples
  \citep{ouchi:03,maier:03,ajiki:03} and filled symbols surveys with
  spectroscopic follow-up
  \citep{cowie:98,kudritzki:00,hu:02,hu:04,kodaira:03,rhoads:03,rhoads:04}.
  The dashed curve in all three panels represents a non-evolving
  luminosity function with $L^{\ast}=3.2\times10^{42}$~erg~s$^{-1}$
  and $\alpha=-1.5$.  The solid curve in the right panel represents
  the number densities probed by our VLT/FORS2 survey; the hatched
  line is our corresponding $1\sigma$ (one-sided) upper limit on the
  LAE number density.  The solid and hatched lines in the middle panel
  represents the same from MS04.  Compared to narrow-band imaging
  searches at $z\sim6.5$, our survey provides better overlap with
  $z\sim3$ surveys at fainter luminosities
  ($L<5\times10^{42}$~erg~s$^{-1}$). The number density of LAE's at
  $z\sim6.5$ is significantly lower than the value suggested by
  \citet{hu:02}'s lensed object, and it seems to be lower than that
  from the unevolved $z\sim3$ luminosity function.
\label{lf}}
\end{figure}

Our multi-slit/filter survey probes flux levels several times fainter
than the narrow-band imaging surveys at $z\sim6.5$ and slightly fainter
than the de-amplified brightness of the lensed HCM6A.  Although we do
not reach the flux levels of the ($z<6$) lensed objects found by
\citet{ellis:01} and \citet{santos:04}, we still expected to find
several LAE's at $z\sim6.5$ brighter than
$L=2\times10^{42}$~ergs~s$^{-1}$, as shown in Fig.~\ref{lf}.  

Our first robust conclusion, as already suggested by the brighter,
wide-angle, narrow-band imaging surveys of \citet{kodaira:03} and LALA
\citep{rhoads:04}, is that the number density of LAE's must be
significantly lower than suggested by the discovery of the lensed
HCM6A; \citet{hu:02} were very fortunate to find this one object.

\begin{figure}
\epsscale{0.7}
\plotone{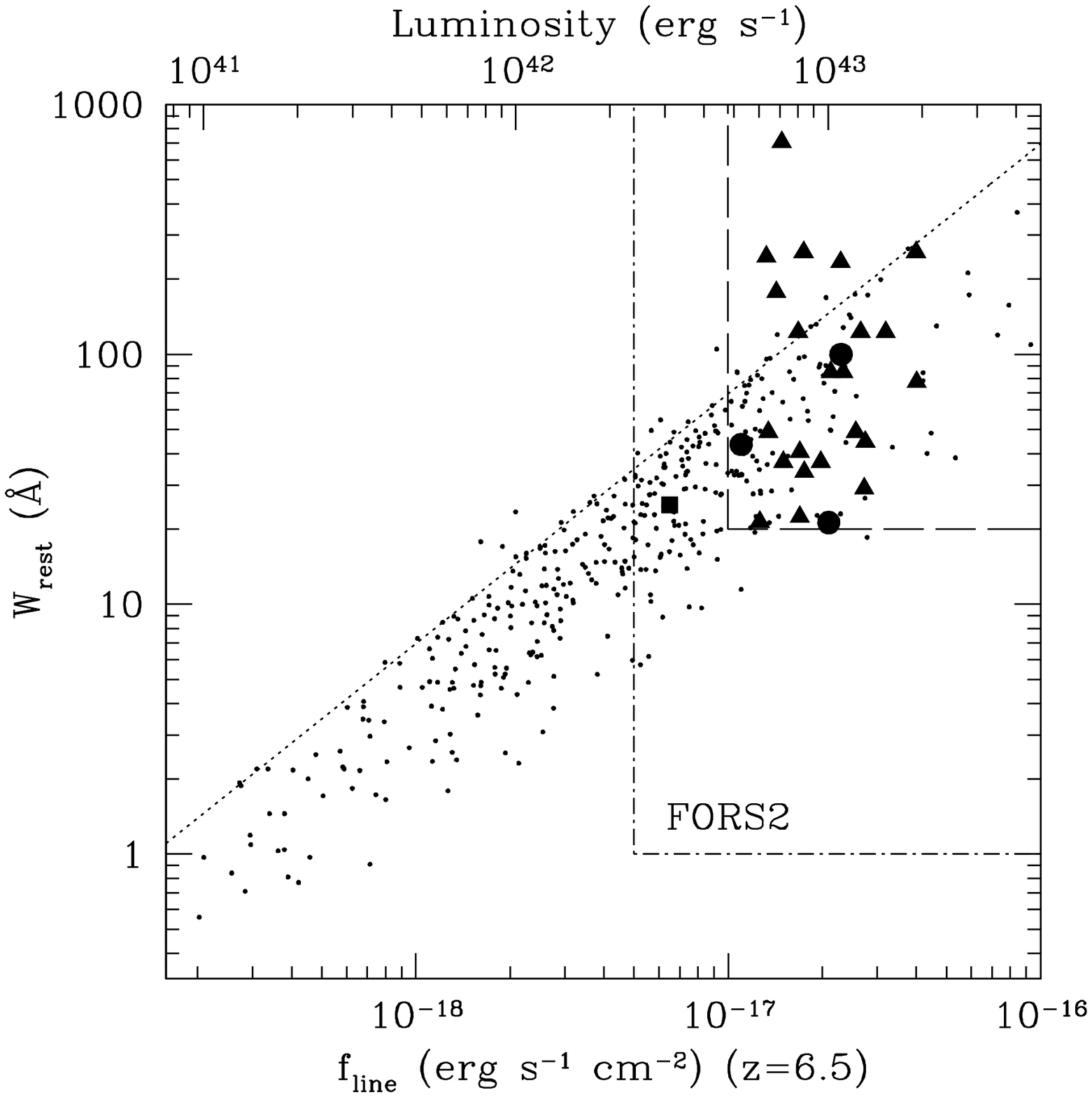}
\caption{Rest Ly$\alpha$ equivalent width versus the flux of the
  Ly$\alpha$ line for the Lyman-break galaxy (LBG) sample at $<z>=3$
  \citep[small circles; ][]{shapley:03} as well as spectroscopically
  confirmed Ly$\alpha$ emitters at $z\sim5.7$ \citep[solid
  triangles;][]{rhoads:03,hu:04} and at $z\sim6.5$ \citep[large solid
  circles;][]{kodaira:03,rhoads:04}; \citet{hu:02}'s de-amplified
  $z=6.56$ LAE is also shown (large solid square).  All the fluxes
  have been redshifted to $z=6.5$.  The long-dashed lines denote the
  nominal limits of narrow-band searches such as LALA while the
  dot-dashed lines denote our FORS2 limits.  The diagonal short-dashed
  line corresponds to a {\it continuum} flux of $I_{AB}=25$; the
  deficit of LBG's above the dashed line is purely due to this
  selection effect.  Here we clearly see the difference between a
  line-selected sample ($e.g.$ vertical and horizontal limits of our
  FORS2 survey) and a continuum selected sample (diagonal dashed
  line). LAE's fainter than a given continuum luminosity can still be
  detected in a line-selected sample such as ours.  Note that although
  dispersed surveys reach much lower equivalent widths than
  narrow-band imaging surveys, they would require even lower flux
  limits to fully exploit this sensitivity given the apparent maximum
  continuum luminosity shown by high redshift galaxies.
\label{eqw_line}}
\end{figure}

\begin{figure}
\epsscale{0.7}
\plotone{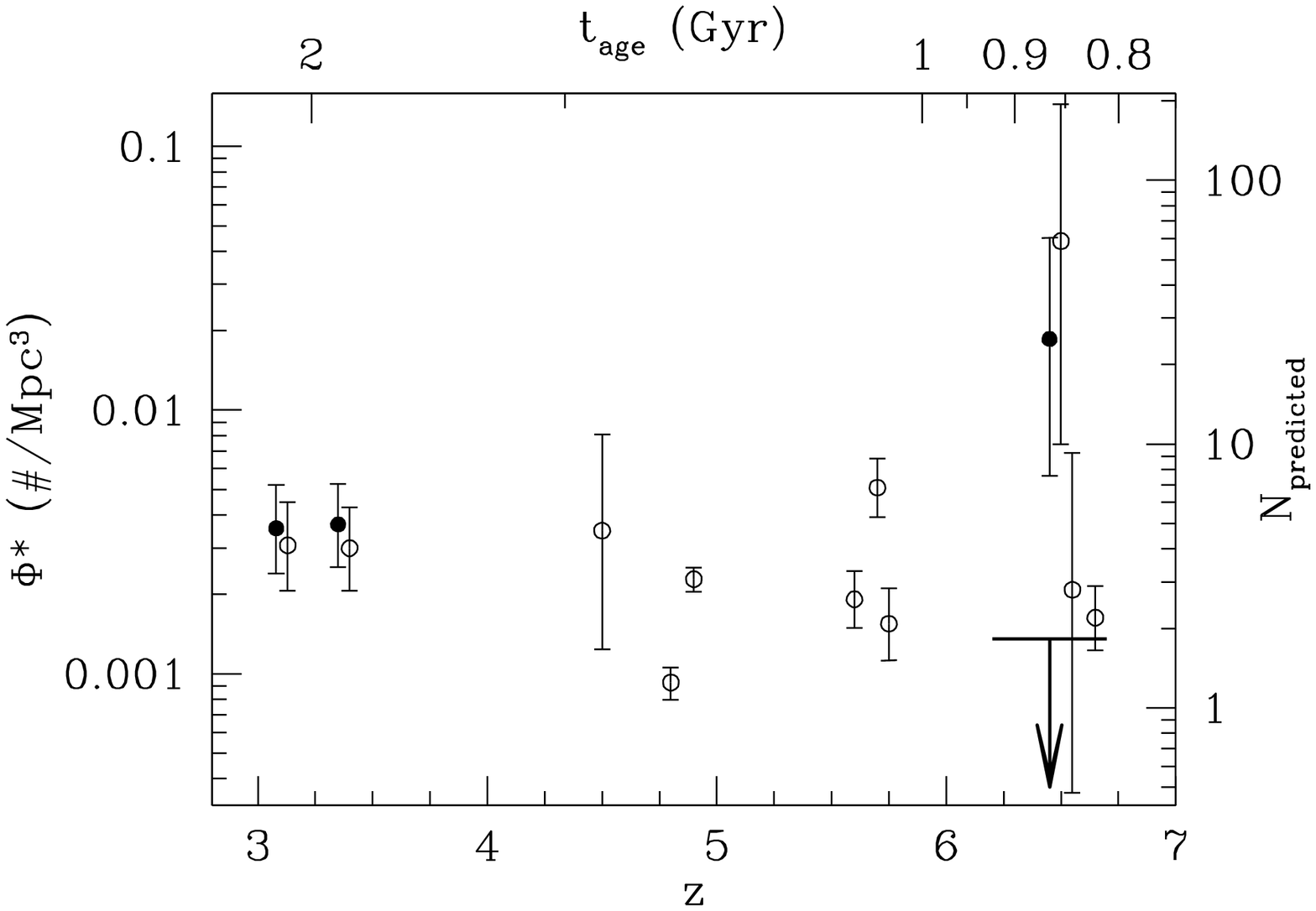}
\caption{Observed number density of Ly$\alpha$ emitters as a function
of redshift.  The solid horizontal line with an arrow denotes our own
84\% (one-sided) confidence level for the number density of LAE's at
$z\sim6.5$.  The solid points for \citet{cowie:98},
\citet{kudritzki:00}, and \citet{hu:02} correspond to the predicted
number of LAE's in our survey volume based on the properties of their
individual objects (see Table~1 for minimum luminosities).  The open
symbols correspond to the predicted number of LAE's from a given
survey estimated by assuming a non-evolving luminosity function with
$L^{\ast}=3.2\times10^{42}$~erg~s$^{-1}$ and $\alpha=-1.5$ (see
Table~1).  The errorbars are estimated from Poisson statistics using
the number of LAE candidates in a given survey, and values for
$N_{pred}$ at a similar redshift are randomly shifted in $z$ for
clarity.  Our results suggest that the number density of LAE's
declines by about a factor of two between $z\sim3$ and $z\sim6.5$.
\label{phi_z}}
\end{figure}

Our null result also suggests that the number density of LAE's with
$L\geq2\times10^{42}$~ergs~s$^{-1}$ is lower at $z\sim6.5$ than at
$z\sim3$.  We stress that compared to narrow-band imaging searches at
$z\sim6.5$, our survey reaches lower flux densities (without lensing)
and thus provides better overlap with $z\sim3$ surveys at fainter
luminosities ($L<5\times10^{42}$~erg~s$^{-1}$; see
Fig.~\ref{eqw_line}).  For $z\sim3$ surveys with flux sensitivities
comparable to ours and spectroscopic confirmation, we use the
properties of the individual objects to determine the number of LAE's
we expected in our own survey, $i.e.$

\begin{equation}
N_{pred}= \sum_{i} \frac{1}{V_i}\times V(L_i)
\end{equation}

\noindent where $V_i$ is the volume occupied by the $i$th object, with
luminosity $L_i$, in the given survey, and $V(L_i)$ is the volume
available to luminosity $L_i$ in our own survey.  The values for
$N_{pred}$ using the properties of the individual LAE's from
\citet{cowie:98} and \citet{kudritzki:00} are shown in Table~1 and
Fig.~\ref{phi_z}.  We also include the same estimate using
\citet{hu:02}'s HCM6A, as this is the only $z>5$ survey with flux
sensitivities comparable to ours.

In additon, we have computed the number of LAE's that we expected to
find based on the major LAE surveys at $3\lesssim z\lesssim6.5$.  In
this case, we adopt a procedure similar to MS04 and assume a Schechter
function with a non-evolving $L^{\ast}=3.2\times10^{42}$ and
$\alpha=1.5$.  We determine $\Phi^{\ast}$ for each survey, and for
these $\Phi^{\ast}$ we then compute the expected number of LAE's
($N_{pred}$) in our own survey.  These results also are listed in
Table~1 and shown in Fig.~\ref{phi_z}.  

Relative to surveys at $z\sim3$, we can exclude with $>95$\% confidence
that the number density stays constant between $z\sim3$ and $z\sim6.5$.

Taken as a whole, the various surveys indicate that the luminosity
function of Ly$\alpha$ emitters is declining in number density around
$L\sim3\times10^{42}$~erg~s$^{-1}$ ($\sim L^{\ast}$; see
Fig.~\ref{lf}).  We estimate the number density of LAE's declines by
$\sim2$ between $z\sim3$ and $z\sim6.5$ (Fig.~\ref{phi_z}).  This
is less than the factor of six observed in Lyman-break galaxy surveys
at $z\sim6$ \citep[e.g.][]{stanway:04}.  However, we note that our
assumed shape for the luminosity function does not fit the LAE surveys
at $z>5$ very well, and that incompleteness in the narrow-band surveys
may become particularly problematic at $L<2L^{\ast}$ (see
Fig.~\ref{lf}).

Finally, our experience with this survey in both phases emphasizes the
difficulty of identifying Ly$\alpha$ from a single emission line.  As
noted by \citet{stern:00} and found in this survey, isolated emission
lines with very large observed equivalent widths ($>200$\AA) are not
exclusively Ly$\alpha$ emission.

\acknowledgements

This research has been supported by the Swiss National Science
Foundation.  We thank A. Shapley and the DLS team for the generous use
of their data, and D. Stern for helpful comments.  We also thank C.
Porciani and P. Norberg for useful discussions.

\bibliographystyle{/home/vy/aastex/apj}
\bibliography{/home/vy/aastex/tran.bib}

\clearpage

\begin{deluxetable}{lrrrr}
\tablecolumns{5}
\tablewidth{0pc}
\tablecaption{Predicted Number of Ly$\alpha$ Emitters\tablenotemark{a}}
\tablehead{
\colhead{Source} & \colhead{$z$} &
\colhead{$L_{min}$ ($10^{42}$)\tablenotemark{b}} &
\colhead{$\Phi^{\ast}$\tablenotemark{c}} &
\colhead{$N_{pred}$\tablenotemark{d}}}
\startdata
\citet{kudritzki:00}&  3.13  &  1.7 & \nodata &  4.8 \\
\citet{cowie:98}    &  3.4   &  2.0 & \nodata &  5.0 \\
\citet{hu:02}       &  6.56  &  2.8 & \nodata & 25.3 \\\\
\hline\\
\citet{kudritzki:00}&  3.13  &  1.7 & 0.0031 &  4.1 \\
\citet{cowie:98}    &  3.4   &  2.0 & 0.0030 &  4.0 \\
\citet{hu:98}       &  4.52  &  4.9 & 0.0035 &  4.7 \\
\citet{shimasaku:04}&  4.79  &  2.6 & 0.0009 &  1.2 \\
\citet{ouchi:03}    &  4.86  &  2.0 & 0.0023 &  3.1 \\
\citet{hu:04}       &  5.7   &  7.1 & 0.0051 &  6.8 \\
\citet{ajiki:03}    &  5.7   &  5.1 & 0.0019 &  2.6 \\
\citet{rhoads:03}   &  5.7   &  6.3 & 0.0015 &  2.1 \\
\citet{rhoads:04}   &  6.5   & 11.1 & 0.0021 &  2.8 \\
\citet{kodaira:03}  &  6.5   &  5.4 & 0.0016 &  2.2 \\
\citet{hu:02}       &  6.56  &  2.8 & 0.0437 & 58.8 \\
\enddata
\tablenotetext{a}{The expected number $N_{pred}$ for the first three
  entries is determined using the properties of the individual Ly$\alpha$
  emitters from the given survey (see text).  These are the only three
  surveys with flux sensitivities comparable to ours and spectroscopic
  confirmation.  For the remaining entries, we assume a luminosity
  function with $L^{\ast}=3.2\times10^{42}$~erg~s$^{-1}$ and
  $\alpha=-1.5$.} 
\tablenotetext{b}{The value we adopted for the minimum luminosity (in
  units of erg~s$^{-1}$) of a given survey to determine $\Phi^{\ast}$.}
\tablenotetext{c}{Number density $\Phi^{\ast}$ in units of 
  Mpc$^{-3}$ ($\Omega_M=0.3$, $\Omega_\Lambda=0.7$, and
  H$_0=70$~km~s$^{-1}$~Mpc$^{-1}$).}     
\tablenotetext{d}{Using the $\Phi^{\ast}$ of the given survey and our
  volume, the number of Ly$\alpha$ emitters ($L\geq2\times10^{42}$) we
  expected to find from assuming a luminosity function (lower
  entries).  Values for the three top entries are determined from Eq.~1.}
\end{deluxetable}

\clearpage

\end{document}